\documentstyle[preprint,aps]{revtex}

\begin{document}
\draft
\title{Electrochemical capacitance of a leaky nano-capacitor}

\author{Xuean Zhao$^1$, Jian Wang$^1$, and Hong Guo$^2$}
\address{$^1$Department of Physics, The University of Hong Kong, 
Pokfulam Road, Hong Kong, China\\
$^2$Centre for the Physics of Materials and Department of Physics,
Mc Gill University, Montreal, PQ, Canda H3A 2T8} 
\maketitle

\begin{abstract}

We report a detailed theoretical investigation on electrochemical 
capacitance of a nanoscale capacitor where there is a DC coupling
between the two conductors. For this ``leaky'' quantum capacitor,
we have derived general analytic expressions of the linear and second 
order nonlinear electrochemical capacitance within 
a first principles quantum theory in the discrete potential 
approximation. Linear and nonlinear capacitance coefficients 
are also derived in a self-consistent manner without the
latter approximation and the self-consistent analysis 
is suitable for numerical calculations. At linear order, the full
quantum formula improves the semiclassical analysis in the tunneling
regime. At nonlinear order which has not been studied before for leaky
capacitors, the nonlinear capacitance and nonlinear nonequilibrium charge
show interesting behavior. Our theory allows the investigation of 
crossover of capacitance from a full quantum to classical regimes
as the distance between the two conductors is changed.

\end{abstract}


\section{Introduction}

The most significant development in electronic devices has been the
progressive miniaturization of them: it is now common to fabricate electron
device structures with dimensions at mesoscopic scale and even at nanoscale. 
One of the important directions in nanoelectronics research is to understand
device properties which relate to the existence of small dimensions. 
In this work, we investigate the notion of electrochemical capacitance
for conductors in the mesoscopic or nanoscale and the nonequilibrium charge 
distribution at the nonlinear level. Using a full first principles quantum
theory, a semiclassical theory, as well as a direct numerical solution, we
construct an overall physical picture on the effects of quantum tunneling 
to the nanoscale capacitance. We also investigate the density of state
correction to capacitance at nonlinear orders of the external bias. 
For a pair of very small conductors, it has been known that the capacitance 
may be quite different from the usual parallel plate capacitance formula 
given by $C_o\sim 1/W$, where $W$ is the distance between the two plates. 
Apart from the usual electrostatic fringe effect, there are quantum 
corrections to the classical formula. Quantum 
corrections come from several sources: a finite density of states of 
the plates, a finite screening length to the electron-electron interactions, 
and quantum tunneling. 

The quantum correction to classical formula due to density of states 
(DOS) has been theoretically\cite{stern,lu} and experimentally\cite{smith} 
investigated in the literature by a number of authors. For semiconductor 
heterojunctions they found that DOS 
contributes a factor to the capacitance given by $C_{DOS} = e^2(dN/dE)$ 
where $dN/dE$ is the total density of states of the system. Thus the total 
capacitance $C$ is a result of a series connection of $C_o$ and $C_{DOS}$: 
$1/C=1/C_o+1/C_{DOS}$. This formula has been theoretically studied from 
a dynamic point of view and was derived in the low frequency limit of an 
AC theory\cite{butphca,but1}. Significantly, these investigations on 
DOS effects focused on the {\it linear} capacitance coefficient $C$, namely
$C$ is the linear coefficient of the charge pile-up on a capacitor plate 
as a function of the external bias voltage. Recently some 
attention\cite{wbg1} has been paid to the nonlinear regime: due to the 
nonlinear bias dependence of local DOS there is also a nonlinear bias
dependence of the nonequilibrium charge. The nonlinear capacitance 
coefficients is one of the topics which will be further investigated 
below.

Mesoscopic electrochemical capacitance has been found\cite{butt2} to obey, 
within a magnetic field, weaker Onsager-Casimir symmetry relations.  
For example it is no longer a symmetric function of magnetic 
field\cite{butt2}. The asymmetry of electrochemical capacitance has been 
observed for a gate close to the edge of a quantum Hall bar\cite{chen}. 
The magneto-electrochemical capacitance of a three-dimensional quantum dot 
with three-probes has been studied numerically in detail in 
Ref.\onlinecite{wei}. It is found that at low magnetic fields the 
magnetocapacitance shows a large asymmetry under a magnetic field 
reversal. At higher fields the capacitance is dominated 
by Aharonov-Bohm type oscillations and the fluctuations of the 
asymmetry is somewhat reduced. For the III-V tunneling heterostructures, 
the contribution of the density of states on the magnetocapacitance is 
also studied\cite{chan}. The investigation of the frequency dependent 
electrochemical capacitance for a parallel plate capacitor within the 
nonequilibrium Green's function formalism show interesting oscillatory 
behavior which is related to the retardation effect of the Maxwell 
equations\cite{wbg}.

As mentioned above, quantum tunneling effect changes the capacitance
value as predicted by the classical formula. This effect was
recently addressed using numerical analysis of atomic 
junctions\cite{jian1}. Numerical calculations\cite{jian1} of aluminum 
atomic junctions with tiny DOS showed that at small distances $W$, the 
electrochemical capacitance $C=C(W)$ actually {\it increases} with $W$ 
which is due to tunneling effect. One expects that at larger $W$
when tunneling effects is diminished, the capacitance would follow 
a crossover to the classical prediction. However due to the very small 
DOS of the atomic junction\cite{jian1}, no crossover to the classical
formula was found in these atomic systems.

The correction to classical capacitance formula due to a finite screening 
length was most clearly demonstrated from a dynamic point of view on the 
electrochemical capacitance, due to the work of Christen and 
B\"uttiker\cite{but2} where a conducting quantum point contact (QPC) was 
found to establish a nonequilibrium charge resulting to a finite 
electrochemical capacitance. In particular they have derived a formula 
for a QPC with a semiclassical method\cite{but2}, 
\begin{equation}
\frac{R}{C}\ =\ \frac{1}{C_{o}}+\frac{1}{e^{2}\frac{dN_{1}}{dE}}+
\frac{1}{e^{2}\frac{dN_{2}}{dE}}\ \ ,
\label{semiclassical}
\end{equation}
where $R$ is essentially a reflection probability of the QPC, 
$C_o$ is geometric capacitance, $dN_1/dE$ and $dN_2/dE$ are the total 
DOS in the regions to the left and to the right of the QPC. 
Qualitatively, the numerical data of the aluminum tunnel 
junction\cite{jian1} were consistent with Eq. (\ref{semiclassical}) 
in that $C$ is proportional to $R$.
Formula (\ref{semiclassical}) is termed ``semiclassical'' because not all
the relevant scattering local partial density of states (LPDOS) were 
included in its derivation. The notion of scattering LPDOS was proposed by 
B\"uttiker\cite{butphca} and subsequently by Gasparian, Christen and 
B\"uttiker\cite{gasp}, and it plays a very important role in low frequency
AC transport as well as nonlinear DC transport. LPDOS describes the
probability of various scattering processes\cite{gasp}. Consider a tunnel
barrier as shown in Fig. (1). An example of a LPDOS is denoted by 
$d\sigma_{22}({\bf r})/dE$ which is the contribution of carriers at position
${\bf r}$ to the DOS, and these carriers come from region $2$ and ultimately 
return to region $2$. Although region $2$ is on the right hand side of the 
tunnel barrier (see Fig. (1)), $d\sigma_{22}({\bf r})/dE\neq 0$ even 
when ${\bf r}$ is on the left hand side of the barrier due to tunneling.
In deriving\cite{but2} Eq. (\ref{semiclassical}) for a QPC, contributions 
such as $d\sigma_{22}({\bf r})/dE$ with position ${\bf r}$ on the other side 
of the QPC, has been neglected. 

In this paper,  we will further investigate nanoscale capacitors where the
two conductors have a DC coupling, namely there is a DC ``leakage'' from one
conductor to the other.  For the linear electrochemical capacitance of
a tunnel barrier, we improve formula (\ref{semiclassical}) by including 
the tunneling contributions of various LPDOS. This way a 
full quantum capacitance formula is derived and will be compared with 
(\ref{semiclassical}). For a single tunnel barrier there is a quantitative
difference between these results in the quantum regime, and 
the difference diminishes as the classical limit is approached. The 
quantum formula and Eq. (\ref{semiclassical}) allow investigations of a 
crossover from tunneling dominated regime to the classical regime, by 
varying the barrier width $W$. Our derivation as well as the derivation 
of Eq. (\ref{semiclassical}) are within the discrete potential 
model\cite{b_euro} that used an approximation where the space is coarse 
grained into a few regions. For the tunnel barrier they are regions to 
the left of the barrier (denoted by $\Omega_1$), to the right of the 
barrier ($\Omega_2$), and the barrier region.  To confirm that this 
approximation does not affect the predictions qualitatively, we have 
carried out extensive numerical calculations of the LPDOS by directly 
solving them without the approximation. 

Recently, the theory of non-linear electrochemical capacitance has been
formulated using the response theory\cite{ma}. The electrochemical 
capacitance of a parallel plate capacitor is a nonlinear function of the
bias voltage due to the finite DOS near the plates as mentioned above.
In this work we will study effect of screening on the nonlinear electrochemical 
capacitance for the ``leaky'' capacitor, which is an important problem 
not investigated before and is relevant for experiments of scanning
capacitance microscopy\cite{exp} applied to nanosystems. We will derive a 
general expression of the second and third order nonlinear quantum 
electrochemical capacitance using the discrete potential model\cite{b_euro}. 
Our analysis naturally deduces, in appropriate approximations, 
a semiclassical expression of the {\it second order} nonlinear 
electrochemical capacitance for QPC. Finally, to compare with results of the 
discrete potential model and semiclassical result, we have directly solved 
the Poisson equation and calculated numerically the linear and the second 
order nonlinear electrochemical capacitance as a function of barrier width 
of a single tunneling barrier.

The main results of our investigation are summarized in the following 
sections. In the next section we present our theory of the nonlinear 
electrochemical capacitance where full quantum tunneling effect is taken 
into account. At the linear order, we compare the quantum formula with the
semiclassical formula; and using scattering Green's functions we derive
second and third order nonlinear results. In Sections III we present 
numerical calculations which is compared with the theoretical analysis.
Finally the last section summarizes the main findings.

\section{Theory}

In general a two-probe system can be considered as having three regions,
a scattering region and two electrodes. This is illustrated in Fig. (1) 
where the scattering region includes the scattering potential barrier, and 
two electrodes are the regions to the left ($\Omega_1$) and to the right 
($\Omega_2$) of the barrier. We are interested in the electrochemical 
capacitance of this system by including the full quantum effects. 
If we refer regions $\Omega_1$ and $\Omega_2$ as the two conductors
of a capacitor, we are dealing with a ``leaky'' capacitor since the 
potential barrier provides a DC coupling between the conductors. 
Far away from the the regions, the system 
is connected to contacts which are viewed as large thermodynamic reservoirs, 
hence in the contacts the electron distributions are Fermi-Dirac.
When a voltage $V_{1}$ is applied at contact 1 and $V_{2}$ at contact 2, 
assume $V_{1}<0$, the electron energy band at contact 1 is changed by 
$d\mu _{1}=eV_{1}$ and at contact 2 by $d\mu_{2}=eV_{2}$. The relative 
electrochemical potential difference is thus $d\mu =d\mu_{1}-d\mu_{2}$: 
due to $d\mu$ electrons are injected into the system. The force acting 
on electrons comes from a combination of external and internal fields. In 
principle, motion of electrons in the total field can be solved by 
Schr\"{o}dinger equation. In particular we will adopt the scattering matrix
approach formulated by Landauer\cite{land}, Imry\cite{imry}, and 
B\"{u}ttiker\cite{buta,butb} to solve the single electron transport problem
which gives the necessary LPDOS needed for the calculation of electrochemical 
capacitance. 

Study of electrochemical capacitance is closely related to the
calculation of changes of the local band $eU({\bf r})$. It is
clear that this local band change near the tunnel barrier is 
different from the shift $d\mu_{k}$ which occurs at the contacts far
away from the barrier. At equilibrium conditions the electron energy 
near the barrier is given by $E_{t}=E+d\mu_{k}-eU({\bf r})$ where 
$E$ is the electron energy at Fermi level without the applied voltage. 
$d\mu_{k}$ denotes the electrochemical potential change in reservoir $k$.
Near the barrier electrons accumulate for regions where $E_{t}>E$ and 
deplete for regions where $E_{t}<E$. It is these accumulated charges which 
we must evaluate. The internal potential build-up $eU({\bf r})$ can be
solved by a self-consistent Poisson equation. For simplicity of discussion,
in the following we use $U_{1}({\bf r})$ and $U_{2}({\bf r})$ to denote 
this potential in regions $\Omega_{1}$ and $\Omega_{2}$ respectively.
Furthermore, analytical derivation of capacitance formula in terms of
microscopic quantities is possible if we use a space-averaged potential 
$U_k$ to replace the space dependent potential $U_k({\bf r})$, as was 
done in Ref.\onlinecite{but2}. This corresponds to the discrete potential 
model proposed by Christen and Buttiker\cite{b_euro}.

We represent the number of electrons in the region $\Omega_{k}$ ($k=1,2$) 
incident from contact $\alpha$ ($\alpha=1,2$) by $\sigma_{k\alpha }$,
which is a function of electron energy $E_{t}$. Hence
$\sigma_{k\alpha}=\sigma_{k\alpha }(E+d\mu_{\alpha }-eU_{k})$. The number of 
electrons without external bias (at equilibrium) is thus 
$\sigma_{k\alpha }(E)$, because $U_{k}\rightarrow 0$ when $d\mu\rightarrow 0$. 
By definition, the electrostatic (geometrical) capacitance $C_o$ between 
the two regions $\Omega_1$ and $\Omega_2$ is given by $C_o=\Delta 
Q_1/(U_1-U_2)$ (or by $C_o=\Delta Q_2/(U_2-U_1)$) where $\Delta Q_k$ ($k=1,2$) 
is the charge measured from the equilibrium value in region $\Omega_k$ 
regardless where they have come from, {\it i.e.}
$\Delta Q_k=\sum_\alpha [\sigma_{k\alpha}(E_t)-\sigma_{k\alpha}(E)]$,
where, to avoid confusion we use $k=I,II$ to denote the regions from now
on. Since there are two electrodes, {\it i.e.} $\alpha=1,2$, $\Delta Q_k$ thus 
consists of two parts. For example, in region $\Omega_I$ ({\it i.e.} $k=I$), 
a part of $\Delta Q_I$ is due to electrons incident from electrode $\alpha=1$ 
which are scattered back to region $k=I$. We denote this part of $\Delta
Q_I$ by $\Delta N_1(\Omega_I)=\sigma_{I1}(E+d\mu_{1}-eU_{1})-\sigma_{I1}(E)$. 
The second part of $\Delta Q_I$ comes from electrons launched at electrode
$\alpha=2$ but ended up in region $k=I$, this part is expressed by
$\Delta N_2(\Omega_{I})=\sigma_{I2}(E+d\mu_{2}-eU_{1})-\sigma_{I2}(E)$.
Hence $\Delta Q_I=\Delta N_1(\Omega_I)+\Delta N_2(\Omega_I)$.

The above partition of local charge according to where it comes from 
can be equally applied to the scattering local partial density of 
states\cite{gasp}. Hence, for example, $d\sigma_{12}(\Omega_I)/dE$ is the LPDOS
which is the DOS for an electron incident from electrode $2$ passing 
through region $\Omega_{I}$ and reaching electrode $1$. Similarly,
$d\sigma_{22}(\Omega_{I})/dE$ is the LPDOS which is the DOS for an 
electron incident from electrode $2$ passing through region $\Omega_{I}$ 
and eventually returning to electrode $2$\cite{foot4}. Both of these LPDOS 
describe 
the tunneling process. This latter term is neglected for a semiclassical 
calculations and is nonzero for a quantum analysis, as emphasized in 
Ref. \onlinecite{gasp}. They both contribute to the electrochemical 
capacitance\cite{foot1} which is the experimentally measured capacitance
defined by
\begin{equation}
\label{cmu}
C_{\mu}\ =\ \frac{eQ_1}{d\mu_1-d\mu_2}\ \ .
\end{equation}
The rest of the paper is devoted to calculate $C_\mu$ including all the
quantum effects discussed above.

Based on the above discussions, we can write down the following two
equations\cite{wbg1} for the classical geometrical capacitance.
Using charges of region $\Omega_I$,
\begin{eqnarray}
C_o\times (U_1-U_2)&=&\sigma_{I1}(E+d\mu_{1}-U_1)-\sigma_{I1}(E) 
\nonumber\\
&&+\sigma_{I2}(E+d\mu_{2}-U_{1})-\sigma_{I2}(E)\ .
\label{c01}
\end{eqnarray}
Using charges of region $\Omega_{II}$,
\begin{eqnarray}
C_o\times (U_2-U_1)&=&\sigma_{II1}(E+d\mu_1-U_2)-\sigma_{II1}(E) 
\nonumber\\
&&+\sigma_{II2}(E+d\mu_2-U_2)-\sigma_{II2}(E)\ .
\label{c02}
\end{eqnarray}
Because the same charge defines electrochemical capacitance $C_\mu$ as
given by Eq. (\ref{cmu}), we have
\begin{equation}
C_o\times (U_1-U_2)=C_\mu\times (d\mu_1-d\mu_2)\ \ .
\label{c03}
\end{equation}
Finally, it is important to remember that the internal electrostatic
potential $U_k$ is a function of the electrochemical potential at the
reservoirs,
\begin{equation}
U_1=U_1(\mu_1,\mu_2)\ ,\ \ \ \ U_2=U_2(\mu_1,\mu_2)\ \ \ .
\label{u1u2}
\end{equation}
In above equations we have set electron charge $e=1$ so that 
$d\mu_\alpha =V_\alpha$ which is the bias voltage at reservoir $\alpha$.

Equations (\ref{c01}, \ref{c02}, \ref{c03}) are the fundamental equations
which we will use to derive quantum corrections to $C_o$ at the linear and
nonlinear orders. Because our theory is gauge invariant, without loss of 
generality we set $V_1=V$ and $V_2=0$ throughout the following analysis.

\subsection{Linear electrochemical capacitance formula}

As discussed above, a semiclassical formula of the linear electrochemical
capacitance has been derived in Ref. \onlinecite{but2} in the form of Eq.
(\ref{semiclassical}). In this subsection we derive a full quantum formula.

Taking derivatives of Eqs. (\ref{c01},\ref{c02}, \ref{c03}) with respect 
to $V$, we obtain
\begin{equation}
C_o(\frac{dU_1}{dV}-\frac{dU_2}{dV})\ =\ \frac{d\sigma_{I1}}{dE_1}
(1-\frac{dU_1}{dV})-\frac{d\sigma_{I2}}{dE_2}\frac{dU_1}{dV} 
\label{du1dv}
\end{equation}

\begin{equation}
C_o(\frac{dU_2}{dV}-\frac{dU_1}{dV})\ =\ \frac{d\sigma_{II1}}{dE_3}
(1-\frac{dU_2}{dV})-\frac{d\sigma_{II2}}{dE_4}\frac{dU_2}{dV} 
\label{du2dv}
\end{equation}

\begin{equation}
C_o(\frac{dU_1}{dV}-\frac{dU_2}{dV})=C_{\mu}
\label{c04}
\end{equation}
where $E_1\equiv E+V_1-U_1$, $E_2\equiv E+V_2-U_1$, $E_3=E+V_1-U_2$, 
and $E_4=E+V_2-U_2$. In deriving the last equation, we have assumed that 
$C_\mu$ has no bias voltage dependence\cite{foot2}. In general the above 
derivatives should be done at a finite bias voltage $V$, but experimentally 
one can control this parameter and use very small voltages\cite{smith} 
$V<<E$. Hence we will evaluate the derivatives at the $V\rightarrow 0$
limit. In the above equations, the quantity $d\sigma_{k\alpha}/{dE_i}$ 
is just the LPDOS in the corresponding regions as discussed above (where we
used the notation such as $d\sigma_{I1}/dE$).

From Eqs. (\ref{du1dv}, \ref{du2dv}, \ref{c04}), eliminating 
$dU_1/dV$ and $dU_2/dV$, we obtain 
\begin{eqnarray}
\frac{\frac{\frac{d\sigma_{I1}}{dE_1}}{\frac{d\sigma_{I1}}{dE_1}+
\frac{d\sigma_{I2}}{dE_2}}-\frac{\frac{d\sigma_{II1}}{dE_3}}
{\frac{d\sigma_{II1}}{dE_3}+\frac{d\sigma_{II2}}{dE_4}}}{C_{\mu }}\ &=&\
\frac{1}{C_o}+\frac{1}{\frac{d\sigma_{II1}}{dE_1}+\frac{d\sigma_{I2}}
{dE_2}} \nonumber \\
&+&\frac{1}{\frac{d\sigma_{II1}}{dE_3}+\frac{d\sigma_{II2}}{dE_4}}
\ \ .
\label{result}
\end{eqnarray}
The electrochemical capacitance $C_\mu$ calculated from this formula is
fully quantum: all the tunneling effects are taken into account through the
appropriate LPDOS which can be evaluated from quantum scattering
calculations (see below).

The general result (\ref{result}) can be reduced to the semiclassical form
Eq. (\ref{semiclassical}) if we apply the semiclassical version of the
LPDOS. In the semiclassical limit, Ref. \onlinecite{but2} showed that
the LPDOS is given by
\begin{equation}
\frac{d\sigma_{k\alpha }}{dE}=\sum_\beta D_{k}[\frac{T}{2}+
\delta_{\beta\alpha}(R\delta_{\beta k}-\frac{T}{2})]
\label{semi1}
\end{equation}
where $T$ is the transmission coefficient, $R$ is related to the
reflection coefficient, $D_I\equiv d\sigma_{I1}/dE_1+d\sigma_{I2}/dE_2$
and $D_{II}\equiv d\sigma_{II1}/dE_1+d\sigma_{II2}/dE_2$ are essentially 
total local DOS in regions $\Omega_I$ and $\Omega_{II}$. 
Substituting (\ref{semi1}) into Eq. (\ref{result}), 
it is straightforward to prove that Eq.(\ref{result}) reduces to the
result of Ref.\onlinecite{but2}:
\begin{equation}
\frac{R}{C_{11}}=\frac{1}{C_o}+\frac{1}{D_I}+\frac{1}{D_{II}}\ \ ,
\label{semi}
\end{equation}
where we used notation $C_{11}$ to denote the {\it linear} electrochemical
capacitance $C_\mu$. If we further set $R=1$, {\it i.e.} no DC coupling 
is allowed between the two regions, formula (\ref{semi}) reduces to the 
familiar electrochemical capacitance of two plates where there is no DC 
current flowing through\cite{but1}.

In Section III we will provide numerical plots of the full quantum and
semiclassical formula, and compare them with direct numerical solution of
the same problem which does not employ the discrete potential model.

\subsection{Nonlinear electrochemical capacitance formula}

We now derive the second order nonlinear electrochemical capacitance from 
the fundamental equations (\ref{c01}), (\ref{c02}), and (\ref{c03}). 
A similar procedure leads to higher order results.
To obtain nonlinear results we expand Eqs.(\ref{c01}) and (\ref{c02}) 
order by order in terms of the bias voltage $V_\beta$ and internal potential
$U_\beta$. The expansion coefficients are energy derivatives of the spectral
function $\sigma_{k\alpha}$, where the first derivative is the linear LPDOS
used in the last section, while the second derivative is the second order
nonlinear LPDOS which is analyzed in the Appendix A. Similarly higher order
derivatives are the corresponding higher order nonlinear LPDOS. 
To simplify notation in the following we denote 
$D_{k\alpha}\equiv d\sigma_{k\alpha}/dE$ and 
$\bar{D}_{k\alpha}\equiv d^2\sigma_{k\alpha}/dE^2$. 

To second order in bias voltage, Eqs.(\ref{c01}) and (\ref{c02}) become
\begin{eqnarray}
C_0(U_1-U_2) &=& \sum_\beta D_{I\beta} V_\beta - D_I U_1 \nonumber \\
&+& \sum_\beta \frac{1}{2} \bar{D}_{I\beta} (V_\beta - U_1)^2
\end{eqnarray}
\begin{eqnarray}
-C_0(U_1-U_2) &=& \sum_\beta D_{II\beta} V_\beta - D_{II} U_2 \nonumber \\
&+& \frac{1}{2} \sum_\beta \bar{D}_{II\beta} (V_\beta - U_2)^2\ \ .
\end{eqnarray}
Using Eq.(\ref{semi1}) and expression (\ref{semi2}) of Appendix A,
in the semiclassical limit the above two equations become
\begin{eqnarray}
& & C_0(U_1-U_2) =D_I(\frac{T}{2}+R)V_1 + D_I \frac{T}{2}V_2 - D_I U_1
\nonumber \\
&+& \frac{1}{2} R \bar{D}_I (V_1-U_1)^2 + \frac{1}{2} T \bar{D}_I
(V_2-U_1)^2
\label{a1}
\end{eqnarray}
and
\begin{eqnarray}
&-&C_0(U_1-U_2) = D_{II}\frac{T}{2}V_1 + D_{II} (\frac{T}{2}+R)V_2 - 
D_{II} U_2 \nonumber \\
&+& \frac{1}{2} T \bar{D}_{II} (V_1-U_2)^2 + \frac{1}{2} R \bar{D}_{II}
(V_2-U_2)^2\ \ .
\label{a2}
\end{eqnarray}

In terms of $C_{11}$ of Eq. (\ref{semi}), we obtain 
internal potential $U_1$ and $U_2$ to first order in voltage,
\begin{equation}
U_1 = RV_1 +\frac{T}{2} (V_1+V_2) -\frac{C_{11}}{D_I} (V_1-V_2)
\label{a3}
\end{equation}
and 
\begin{equation}
U_2 = RV_2 +\frac{T}{2} (V_1+V_2) +\frac{C_{11}}{D_{II}} (V_1-V_2)\ \ .
\label{a4}
\end{equation}
Substituting Eqs.(\ref{a3}) and (\ref{a4}) into the quadratic terms of
Eqs.(\ref{a1}) and (\ref{a2}), we obtain
\begin{eqnarray}
C_0(U_1-U_2) &=& D_I(\frac{T}{2}+R)V_1 + D_I \frac{T}{2}V_2 - D_I U_1
\nonumber \\
&+& \frac{1}{2} R \bar{D}_I (\frac{T}{2} + \frac{C_{11}}{D_I})^2
(V_1-V_2)^2 
\nonumber \\
&+& \frac{1}{2} T \bar{D}_I
(R+\frac{T}{2}-\frac{C_{11}}{D_I})^2 (V_1-V_2)^2
\end{eqnarray}
and
\begin{eqnarray}
-C_0(U_1-U_2) &=& D_{II}\frac{T}{2}V_1 + D_{II} (\frac{T}{2}+R)V_2 - D_{II} 
U_2 \nonumber \\
&+& \frac{1}{2} T \bar{D}_{II}(R+\frac{T}{2}-\frac{C_{11}}{D_{II}})^2 
(V_1-V_2)^2 \nonumber \\
&+& \frac{1}{2} R \bar{D}_{II} (\frac{T}{2}+\frac{C_{11}}{D_{II}})^2 
(V_1-V_2)^2\ .
\end{eqnarray}

Combining the above two equations, we finally arrive at
\begin{equation}
C_0(U_1-U_2)= C_{11} (V_1-V_2) +\frac{1}{2} C_{111} (V_1-V_2)^2
\label{second1}
\end{equation}
with the nonlinear capacitance 
\begin{eqnarray}
C_{111} &=& C_{11} \left[ \frac{\bar{D}_I}{D_I} 
(\frac{T}{2}+\frac{C_{11}}{D_I})^2 - \frac{\bar{D}_{II}}{D_{II}} 
(\frac{T}{2}+\frac{C_{11}}{D_{II}})^2 \right] \nonumber \\
&+& \frac{T}{R} C_{11} \left[
\frac{\bar{D}_I}{D_I} (R+\frac{T}{2}-\frac{C_{11}}{D_I})^2 \right.
\nonumber \\
& & \left. -\frac{\bar{D}_{II}}{D_{II}} (R+\frac{T}{2}-
\frac{C_{11}}{D_{II}})^2 \right]
\label{general}
\end{eqnarray}
This result indicates that the second order nonlinear electrochemical
capacitance can be expressed in terms of microscopic quantities such
as the various LPDOS as well as transmission and reflection coefficients.
All of these are calculable and have been studied before. Hence this 
result is very useful in practical predictions of nonlinear capacitance 
coefficient, and it is valid even if there is a DC coupling between 
the two polarization regions of the conductor. 

The general expression (\ref{general}) is reduced in certain limiting 
situations. First, for a spatially symmetric system where $D_I=D_{II}$ and 
$\bar{D}_I=\bar{D}_{II}$, Eq. (\ref{general}) gives $C_{111}=0$. This is 
expected due to symmetry: 
since $C_{111}$ is the coefficient of the charge distribution expanded in 
second order of bias voltage ({\it e.g.} Eq. (\ref{second1})), it must 
vanish as charge $Q \rightarrow -Q$ for symmetrical systems when 
$V\rightarrow -V$.
Second, for a capacitor without DC coupling between the two conductors,
{\it i.e.} for cases $T=0$ identically, Eq.(\ref{general}) becomes
\begin{equation}
C_{111} = C_{11}^3 (\frac{\bar{D}_I}{D_I^3}- \frac{\bar{D}_{II}}{D_{II}^3})
\label{non}
\end{equation}
which was first derived in a response theory\cite{ma}. Finally, 
a point worthy some discussion is the ``resonant transmission point'' 
by setting $T=1$ and $R=0$. For this case from Eq. (\ref{semi}) the 
linear electrochemical capacitance $C_{11}=0$. But from Eq. (\ref{general}) 
$C_{111}\neq 0$ and is given by
\begin{equation}
C_{111} = \frac{1}{4}(\frac{\bar{D}_I}{D_I}-\frac{\bar{D}_{II}}{D_{II}})
\frac{1}{C_0^{-1}+D_I^{-1}+D_{II}^{-1}}\ \ ,
\label{c111}
\end{equation}
which is generally nonzero. Apparently we would expect no charge 
accumulation when $T=1$ hence $C_{111}$ and all other capacitance
coefficients would vanish. However the $T=1$ limit in the above formula 
only states the fact that {\it injected} charges are going through from 
one capacitor plate to the other at the {\it linear} order, and it does 
not implicate the behavior of the charges at nonlinear order where
in general $T=T(E,U)$. Thus in setting $T(E)=1$ in Eq. (\ref{general})
is not the true resonant transmission point: at nonlinear order the resonance
occurs at $T(E,U)=1$.

\subsection{Analysis beyond discrete potential model}

So far we have derived the linear and nonlinear electrochemical 
capacitance coefficients within the discrete potential model, 
in which the internal potential $U_k$ is parametrized in terms of a 
geometrical capacitance $C_o$. This parametrization is necessary in 
order to carried out analytical derivations, and it is adequate to 
reveal qualitative features of the physics. On the other hand, 
if one is willing to perform numerical calculations, it is possible 
to go beyond the discrete potential approximation. In this case we 
can solve the internal potential $U=U({\bf r})$ from a self-consistent 
Poisson equation. In this subsection we derive capacitance expressions
which are suitable beyond the discrete potential model.

We start from the charge pile-up written as a three-dimensional 
spatial integral of the charge density\cite{ma} 
\begin{eqnarray}
Q_\alpha &=& \int_{\Omega_\alpha} \rho(x) d^3x  \nonumber \\
&\equiv& \sum_\beta C_{\alpha
\beta} V_\beta + \frac{1}{2} \sum_{\beta \gamma} C_{\alpha \beta \gamma}
V_\beta V_\gamma +...\ \ \ .
\label{cc}
\end{eqnarray}
Ref. \onlinecite{wbg1} has shown that charge density $\rho (x)$ 
is given in terms of the linear and nonlinear LPDOS, as 
\begin{eqnarray}
\rho(x) &=& \sum_\alpha \frac{d\sigma_\alpha(x)}{dE} (V_\alpha -U(x)) 
\nonumber \\
&+& \frac{1}{2} \sum_\alpha \frac{d^2\sigma_\alpha(x)}{dE^2} 
(V_\alpha - U(x))^2 + ...\ \ \ .
\label{qq}
\end{eqnarray}

To proceed further we must solve the internal Coulomb potential $U(x)$ 
by the Poisson equation
\begin{equation}
-\nabla^2 U(x) = 4\pi \rho(x)\ \ \ .
\label{poisson}
\end{equation}
As done previously\cite{but1,ma}, for perturbative analysis of the  
electrochemical capacitance we introduce the characteristic 
potential $u(x)$
\begin{equation}
U(x) = \sum_\alpha u_\alpha(x) V_\alpha +\frac{1}{2} \sum_{\alpha \beta }
u_{\alpha \beta} V_\alpha V_\beta + ...\ \ \ .
\label{char}
\end{equation}
Hence instead of solving $U(x)$ we solve for $u(x)$ order by order.
From Eqs.(\ref{qq}) to (\ref{char}), we find Poisson like equations 
for the characteristic potentials\cite{foot3}
\begin{equation}
-\nabla^2 u_\alpha +4\pi \frac{d\sigma}{dE} u_\alpha = 4\pi 
\frac{d\sigma_\alpha}{dE}
\label{u1}
\end{equation}
and 
\begin{equation}
-\nabla^2 u_{\alpha \beta} +4\pi \frac{d\sigma}{dE} u_{\alpha \beta} = 
4\pi \frac{d\tilde{\sigma}_{\alpha \beta}}{dE}
\label{u11}
\end{equation}
where\cite{ma}
\begin{equation}
\frac{d\tilde{\sigma}_{\alpha \beta}}{dE} = \frac{d^2\sigma_\alpha}{dE^2} 
\delta_{\alpha \beta} -\frac{d^2\sigma_\alpha}{dE^2} u_\beta - 
\frac{d^2\sigma_\beta}{dE^2} u_\alpha +\frac{d^2\sigma}{dE^2} u_\alpha u_\beta
\end{equation}

With the help of Eqs.(\ref{u1}) and (\ref{u11}), the electrochemical 
capacitance can be calculated from the following expressions,
\begin{equation}
C_{\alpha \beta} \equiv \int_{\Omega_\alpha} Q_\beta(x) d^3 x
= \int_{\Omega_\alpha} \left(\frac{d\sigma_\beta}{dE} - 
\frac{d\sigma}{dE} u_\beta\right) d^3x
\label{cc11}
\end{equation}
\begin{equation}
C_{\alpha \beta \gamma} \equiv \int_{\Omega_\alpha} Q_{\beta \gamma}(x) d^3 x
= \int_{\Omega_\alpha} \left(\frac{d\tilde{\sigma}_{
\beta \gamma}}{dE} - \frac{d\sigma}{dE} u_{\beta \gamma}\right) d^3x
\label{cc111}
\end{equation}
where $Q_\beta(x)$ and $Q_{\beta \gamma}(x)$ are linear and nonlinear
nonequilibrium charge distributions. These results are useful for 
numerical calculations where all the quantities on the right hand side
can be obtained accurately. For instance Eq. (\ref{cc11}) has been used
in the analysis of atomic junctions\cite{jian1}. Eq. (\ref{cc111})
is derived for the first time here.

To end this section we note that in a numerical calculation, the 
LPDOS $d\sigma_\alpha/dE$ can be calculated using the scattering 
wavefunction\cite{but3} 
\begin{equation}
\frac{d\sigma_\alpha(x)}{dE} = \frac{1}{hv} |\psi(x)|^2
\label{injec}
\end{equation}
where $v$ is the velocity of the carrier and $\psi(x)$ is the scattering 
wavefunction for incident wave coming from lead $\alpha$. Eqns. 
(\ref{u1}, \ref{u11}) can be numerically solved on a three-dimensional
grid,  for instance a multi-grid technique was employed in Ref. 
(\onlinecite{jian1}) for such a purpose.

\section{Numerical Results}

In this section we present numerical plots for electrochemical
capacitance of the tunnel barrier in Fig. (1). The numerical 
curves were obtained along two lines: by plotting the analytical 
expressions (\ref{result}, \ref{semi}, \ref{general}) which 
are within the discrete potential model; and by direct numerical 
solution of the self-consistent internal potential $U({\bf r})$ 
and then applying expressions (\ref{cc11}) and (\ref{cc111}).

To be specific, we choose a numerical calculation box with size 
$x_{L}-x_{R}=12\lambda_F$ where $\lambda_F$ is Fermi wavelength of the
scattering electron. Here $x_{L,R}$ are the positions of left and right 
boundary of the calculation box. We fix the tunnel barrier of width $W$
symmetrically in the center of the calculation box. This way the regions
$\Omega_I$ and $\Omega_{II}$ discussed above are given by the space between
the calculation box and the barrier walls. The quantum scattering problem
by this single barrier is easily solved, from which we obtain
various LPDOS using the scattering wavefunction according to 
Eq.(\ref{injec}). To apply expressions (\ref{result}, \ref{semi}, 
\ref{general}) which are appropriate for the discrete potential model, 
we spatially average these LPDOS in the respective regions which gives 
us the corresponding quantities in these expressions.
On the other hand, in applying expressions (\ref{cc11}, \ref{cc111}) which
uses the full spatial dependent internal potential, the spatial integration 
range should be the Thomas-Fermi screening length\cite{but1} as discussed 
in Appendix B. The screening length is determined\cite{but1} by solving 
the Poisson-like equations (\ref{u1}) (and (\ref{u11}) in the nonlinear
case). From now on we will use atomic units such that $\hbar =2m=e=1$. In 
typical nanoscale systems\cite{sohn} with charge density $10^{15}$,
Fermi wavelength $\lambda_F \sim 47 nm$. In the following we use 
$\lambda_F$ to set the unit for length and choose Fermi energy $E_F=0.31$.

Fig.(2) plots the linear electrochemical capacitance obtained from
different approaches as a function of the barrier width $W$ for the
fixed barrier height $H_0=0.8$: (a). the electrochemical capacitance $C$ 
calculated numerically from Eq.(\ref{cc11}) (solid line); (b). the analytic 
quantum electrochemical capacitance formula in the discrete potential 
approximation $C_{q}$ from Eq.(\ref{result}) (dotted line); (c). 
the semiclassical electrochemical capacitance\cite{but2} $C_{s}$ of 
Eq. (\ref{semi}) (dashed line); (d). the effective classical geometric 
capacitance $C_o\sim 1/W$ (dot-dashed line). For very large the barrier 
width, it is clearly shown that all curves approach to the classical 
behavior $\sim 1/W$ since quantum tunneling effect is negligible. For 
thin barriers where tunneling effect is significant, the behavior of 
electrochemical capacitances $C$, $C_{q}$, and $C_{s}$ are completely 
different from the classical regime. In this quantum regime as one 
increases the barrier width, the electrochemical capacitance increases 
rather than decreases. This increasing behavior at very small $W$ is 
expected since tunneling tends to diminish charge polarization, thus 
$C\sim 0$ when $W\sim 0$. Hence $C(W)$ should indeed start from small
values and increase a bit before it goes down when $W$ is large enough. 

To examine the DOS correction which is another quantum effect,
we note that one can only separate out the geometrical effect from the 
DOS effect in the semiclassical limit (as in Eq. (\ref{semi})), and 
in general these effects are mixed. Furthermore, in a discrete potential 
model all the quantities (both in quantum and semiclassical calculations) 
are {\it spatially averaged}, hence capacitances are under-estimated. 
This is why both $C_q$ and $C_s$ curves are consistently below the 
full numerical solution $C$. Fig.(2) shows some difference between the 
quantum result $C_q$ and semiclassical result $C_s$. To understand this 
difference we have plotted the partial DOS $dn_{11}(\Omega_{II})/dE$ 
(solid line) and $dn_{12}(\Omega_{II})/dE$ (dotted line) in the inset of
Fig.(2). As expected, $dn_{11}(\Omega_{II})/dE$ goes to zero for large 
barrier widthes where the semiclassical theory is a good approximation.
It is nonzero in the quantum tunneling regime for small barrier width. 
$dn_{11}(\Omega_{II})/dE$ is also numerically much less than 
$dn_{12} (\Omega_{II})/dE$. Hence neglecting $dn_{11}(\Omega_{II})/dE$ in 
the semiclassical analysis gives a small difference between $C_s$ and 
$C_q$ in the tunneling regime (see Fig. (2)). To further compare with 
the semiclassical result of QPC of Ref.\onlinecite{but2}, we have also 
examined the behavior of capacitance by varying the barrier height $H_0$ 
for a fixed barrier width $W$: the results using Eqs. 
(\ref{cc11},\ref{result},\ref{semi}) are, again, similar in the quantum regime
and the same in the classical regime. When the barrier height $H_{0}$ 
is relatively small, the appearance of quantum mechanism leads to a 
correction for semiclassical electrochemical capacitance. 

The physical behavior of second order nonlinear electrochemical capacitance
coefficient $C_{111}$ can be studied for an asymmetric barrier: as
discussed above $C_{111}=0$ for symmetric systems (see Eq.(\ref{general})). 
For the asymmetric barrier where the barrier heights are $H_1=0.2$ and 
$H_2=1.0$, shown in the inset of Fig.(3a), Fig. (3a) plots $C_{11}$ versus 
$W$ and Fig.(3b) plots $C_{111}$. The linear coefficient $C_{11}$ 
is very similar to that of Fig.(2) of a symmetric barrier.
For the full quantum numerical result (solid line), $C_{111}$ starts from 
zero and becomes {\it negative} for very thin barrier, reaches minimum at
around $W=1.0$, and then oscillates around zero. The oscillatory 
behavior can be traced to oscillations in second order DOS 
$\bar{D}=d^2N/dE^2$ of Eq.(\ref{general}). In
the inset of Fig.(3b), we plot PDOS $D_I$ and $\bar{D}_I$.
Correlating the behavior of $C_{111}$ and PDOS, it is clear that the 
negative values of $C_{111}$ is due to $\bar{D}$. 

In Fig.(4), the linear and nonlinear nonequilibrium charge distribution 
for this asymmetric barrier, $Q_1(x)$ and $Q_{11}(x)$, are plotted. 
These quantities, especially $Q_{11}(x)$, have not been studied 
carefully before. It is thus interesting to offer several observations. 
(a) The linear charge distribution $Q_1(x)$ is in the form of a resistance 
dipole\cite{datta}, whereas the nonlinear charge $Q_{11}(x)$ is more like 
a quadrupole. (b). The linear charge distribution is numerically much 
larger than the nonlinear charge distribution. The total charges are 
conserved, {\it i.e.}, $\int Q_1(x) dx=\int Q_{11}(x) dx = 0$.
(c). In the discrete potential model, the average nonlinear charge 
$Q_{11}$ is numerically even smaller. Because of this spatial average, 
the nonlinear charge distribution becomes a dipole in the discrete 
potential model. This is responsible for the difference between 
full quantum calculation and that of the discrete potential model. 

\section{Discussion and Summary}

In this work we have investigated the quantum version of a ``leaky
capacitor'' in the coherent nanoscale regime in both linear and nonlinear
order in terms of the external bias voltage. We have derived an 
analytic formula of electrochemical capacitance where the two plates have
a DC coupling, and tunneling effects between the two plates are fully
included by explicitly using partial local density of states 
$dn_{\alpha \alpha} (\Omega_k)/dE$. Within the discrete potential model 
where all quantities are averaged over the polarization regions, 
analytic expressions for the linear and second order 
nonlinear electrochemical capacitance have been derived. In addition,
linear and nonlinear capacitance coefficients formula are derived in terms
of the self-consistent potential, and these formula are suitable for
numerical calculations in the full quantum regime. Our calculation showed
that the analytic results using the discrete potential model are consistent
with the full numerical solution, for the single tunnel barrier structure.
The electrochemical capacitance formula derived in this paper are suitable
for analyzing AC transport at relatively low frequency.  At very high frequency,
one has to consider radiation effect thus the quantum equation must be solved
self-consistently with the full Maxwell equation instead of the 
Poisson equation used here.

Quantum behavior of the electrochemical capacitance is found in the 
tunneling regime that the capacitance increases with the barrier width 
$W$. This is in clear contrast to the classical behavior of $1/W$.
What is the condition that this non-classical phenomenon be observed ?
Let's consider this question using the semiclassical formula\cite{but2}
Eq. (\ref{semi}) which can be rewritten as
\begin{equation}
C\approx \frac{R}{\frac{1}{C_o}+\frac{1}{D}}\ \ .
\label{eq1}
\end{equation}
For tunneling, $R\approx [1 - \exp(-W/l)]$ where $l$ is a characteristic 
length depending on system details such as the barrier heights. 
When $C=C(W)$ increases with $W$, we have $\partial C/\partial W >0$ which
gives to a condition for the range of $W$ that allows the nonclassical 
behavior. Using Eq. (\ref{eq1}), for tiny capacitor plates $D<<C_o$, 
one can have a reasonable and experimentally accessible range of order 
$l$. On other hand for large plates $D >> C_o$, one can not observe the 
nonclassical effect unless $W$ is several orders smaller than $l$ which is
not experimentally accessible. Hence the condition to observe the
non-classical behavior is tunneling and also small DOS. Systems which 
satisfy these conditions are nanoscale capacitors, whereas capacitors 
with large plates such as Josephson junctions (macroscopic) do not satisfy 
the DOS condition.

The nonlinear theory developed here can be pushed to higher order
analytically within the discrete potential model. At linear order the full
quantum formula (\ref{result}) and the semiclassical formula (\ref{semi})
give certain numerical difference in the quantum regime. There is a more
substantial numerical difference between the discrete potential model and
the full self-consistent numerical calculation using expressions
(\ref{cc11}) and (\ref{cc111}), although all these results are qualitatively
consistent. At second nonlinear order, the nonequilibrium charge distribution
behaves as a quadrupole, this is compared to the resistance dipole when
linear order charge is considered. It is interesting to note that the
nonlinear charge can be nonzero when linear charge is zero: this happens at
the linear resonance point. The nonlinear capacitance coefficient also
behaves quite differently from the its linear counterpart, as shown by its
oscillatory behavior linked to the nonlinear LPDOS.

{\bf Acknowledgments.}
We gratefully acknowledge the financial support by a RGC grant from the SAR 
Government of Hong Kong under Grant Number HKU 7115/98P.
H.G. is supported by NSERC of Canada and FCAR of Qu\'ebec. 
JW would like to thank National Center for Theoretical Physics for 
hospitality where part of this work was completed. ZXA thanks Professor 
M. B\"uttiker for a helpful communication concerning the physics of 
electrochemical capacitance. We thank Computer Center of The 
University of Hong Kong for computer facilities.

\section{Appendix A.}

The nonlinear electrochemical capacitance depends on the nonlinear LPDOS, as
shown in Section II. In this Appendix, we derive the nonlinear (2nd order)
LPDOS using Green's functions.  In particular we have to relate the second 
order LPDOS $d^2\sigma_{\alpha \beta}/dE^2$ to the total LDOS 
$d^2\sigma/dE^2$, where indices $\alpha$, $\beta$ label the leads. 

We start from the definition of LPDOS expressed in terms of the Green's 
function\cite{butt1},
\begin{equation}
\frac{d\sigma_{12}(x)}{dE} = Re[2\pi i (G^r \Gamma_2 G^a \Gamma_1 G^r)_{xx}]
\label{e1}
\end{equation}
where $G^r$ is the retarded Green's function, $\Gamma_\alpha$ is the 
linewidth function, and $Re[...]$ denotes the real part of $[...]$. 
Using relation\cite{gasp} 
\begin{equation}
G^r_{x x_1} G^r_{x_2 x} = G^r_{xx} G^r_{x_2 x_1}\ \ \ ,
\label{gas1}
\end{equation}
we have 
\begin{eqnarray}
(G^r M G^r)_{xx} &=& \sum_{x_1 x_2} G^r_{x x_1} M_{x_1 x_2} G^r_{x_2 x}
\nonumber \\
&=& G^r_{xx} \sum_{x_1 x_2} G^r_{x_2 x_1} M_{x_1 x_2} \nonumber \\
&=& G^r_{xx} Tr[G^r M]
\label{relation}
\end{eqnarray}
where $M$ is a matrix. Eq.(\ref{e1}) becomes,
\begin{eqnarray}
\frac{d\sigma_{12}}{dE} &=& -2\pi Im[G^r_{xx} Tr(\Gamma_2 G^a \Gamma_1 G^r)]
\nonumber \\
&=& \frac{i}{2\pi} (G^r_{xx} - G^a_{xx}) T = \frac{T}{2} \frac{d\sigma}{dE}
\label{dn12}
\end{eqnarray}
where $T = Tr(\Gamma_2 G^a \Gamma_1 G^r)/(4\pi^2)$ is the transmission 
coefficient\cite{datta} which is a real quantity; $d\sigma/dE$ is the 
total local DOS. 

Taking energy derivative of Eq.(\ref{e1}), we obtain
\begin{eqnarray}
\frac{d^2\sigma_{12}}{dE^2} &=& 2\pi Im[(G^r G^r \Gamma_2 G^a \Gamma_1 
G^r)_{xx} \nonumber \\
&+& (G^r \Gamma_2 G^a G^a \Gamma_1 G^r)_{xx} \nonumber \\
&+& (G^r \Gamma_2 G^a \Gamma_1 G^r G^r)_{xx}]\ \ .
\label{e2}
\end{eqnarray}
The first term of (\ref{e2}) can be simplified using Eq.(\ref{relation})
as follows,
\begin{eqnarray}
& &(G^r G^r \Gamma_2 G^a \Gamma_1 G^r)_{xx} = G^r_{xx} Tr (G^r \Gamma_2 G^a 
\Gamma_1 G^r) \nonumber \\
& =& G^r_{xx} Tr(G^r) Tr(\Gamma_2 G^a \Gamma_1 G^r) = \frac{T}{4\pi^2} 
(G^r G^r)_{xx}\ .
\end{eqnarray}
The other two terms of Eq.(\ref{e2}) can be simplified in a 
similar fashion. We thus have
\begin{eqnarray}
\frac{d^2\sigma_{12}}{dE^2} &= & \frac{T}{2\pi} Im[2 (G^rG^r)_{xx} + 
(G^r G^a)_{xx}] \nonumber \\
&=&-\frac{i T}{\pi} [(G^rG^r)_{xx}-(G^a G^a)_{xx}] = T \frac{d^2\sigma}{dE^2}
\ .
\label{d2n12}
\end{eqnarray}
In deriving the last equation we used the fact that $G^r G^a$ is 
a real quantity. Using (\ref{d2n12}) we find
\begin{eqnarray}
\frac{d^2\sigma_{11}}{dE^2} + \frac{d^2\sigma_{22}}{dE^2} &=& 
\frac{d^2\sigma}{dE^2} - \frac{d^2\sigma_{12}}{dE^2} - 
\frac{d^2\sigma_{21}}{dE^2} \nonumber \\
&=& (1-2T) \frac{d^2\sigma}{dE^2}\ \ .
\label{dn11}
\end{eqnarray}

Now we consider a system with a DC current passing through. Due to 
polarization we again divide the system into two regions $\Omega_I$ and
$\Omega_{II}$. In the semiclassical treatment where the tunneling is neglected, 
the partial DOS $d^2\sigma_{22}/dE^2=0$ in region I, and similarly 
$d^2\sigma_{11}/dE^2=0$ in region II. These relations and 
Eqs.(\ref{d2n12}) and (\ref{dn11}) lead to
\begin{equation}
\frac{d^2\sigma_{k \alpha \beta}}{dE^2} = \frac{d^2\sigma}{dE^2} \{ T 
+\delta_{\alpha \beta} [(1-2T) \delta_{\alpha k} -T]\}
\end{equation}
where $k$ labels the polarization region $\Omega_k$. For two probe system, 
it gives
\begin{equation}
\frac{d^2\sigma_{k \alpha}}{dE^2} = \frac{d^2\sigma}{dE^2}[T+(1-2T) 
\delta_{\alpha k}]\ \ .
\label{semi2}
\end{equation}
This expression is the semiclassical second order nonlinear LPDOS, 
which is in contrast to the semiclassical linear LPDOS Eq.(\ref{semi1}) 
derived in Ref. \onlinecite{but2}. The nonlinear LPDOS plays a crucial 
role in determining the nonlinear electrochemical capacitance, as given 
in Section II. 

\section{Appendix B}

In this appendix, we give an example of calculating the second order 
nonlinear capacitance $C_{111}$ by directly solving Poisson equation. 
This can be done analytically only for very simple systems. 

Consider a system which consists of three regions: two leads 
(region I and III) and an infinite potential barrier (region II). 
Since the calculation is perturbative, we have to calculate the linear 
characteristic potential by solving Poisson equation Eq.(\ref{u1}). We 
assume that the partial local DOS $d\sigma_1/dE$ and $d^2\sigma_1/dE^2$ 
are constant in region I and zero otherwise\cite{but1}. Similarly 
$d\sigma_2/dE$ and $d^2\sigma_2/dE^2$ are constant in region III and zero 
otherwise. The solution of the Poisson equation Eq.(\ref{u1}) is

\begin{eqnarray}
& &~~{\it region ~ I:} ~~~  u_1 = 1 - A_1 \exp(\frac{x}{\lambda_1}) \nonumber \\
& &~{\it region ~ II:} ~~~  u_1 = a_1 x +b_1  \nonumber \\
& &{\it region ~ III:} ~~~  u_1 = A_2 \exp(-\frac{x}{\lambda_2})
\label{sol}
\end{eqnarray}
where $A_1$, $A_2$, $a_1$, and $b_1$ are constants to be determined. In 
Eq.(\ref{sol}), we have defined the screening length $\lambda_\alpha^{-2} =
4\pi d\sigma_\alpha/dE$ and the boundary conditions\cite{but1} that $u_1 
\rightarrow 1$
as $x \rightarrow -\infty$ and $u_1 \rightarrow 0$ as $x \rightarrow \infty$
have been used. Using the boundary condition that $u_1$ and $du_1/dx$ be 
continuous at $x=a/2$ and $-a/2$, it is straightforward to find

\begin{eqnarray}
a_1&=&\frac{1}{a+\lambda_1+\lambda_2}~,~~~ b_1= \frac{a_1}{2}(a+2\lambda_2)
\nonumber \\
A_1 &=& a_1 \lambda_1 \exp(\frac{a}{2\lambda_1})~, ~~~
A_2 = a_1 \lambda_2 \exp(\frac{a}{2\lambda_2})
\end{eqnarray}
The linear electrochemical capacitance can be obtained immediately,
\begin{eqnarray}
C_{11} &\equiv& \int_{\Omega_I} \frac{\partial \rho(x)}{\partial V_1} dx 
\nonumber \\
&=& \frac{-1}{4\pi} \int_{\Omega_I} \nabla^2 u_{1} dx \nonumber \\
&=& \frac{-1}{4\pi} \nabla u_{1}|_{-a/2} \cdot A \nonumber \\
&=& \frac{A}{4\pi} \frac{1}{a+\lambda_1+\lambda_2}
\end{eqnarray}
where $A$ is the cross-section area of the metallic wire. 
Using the global DOS $dN_\alpha/dE = {\it Volume} ~ d\sigma_\alpha/dE = 
\lambda_\alpha A d\sigma_\alpha/dE = A/4\pi \lambda_\alpha$, we arrive 
at the result first obtained by B\"uttiker\cite{but1},

\begin{equation}
\frac{1}{C_{11}} = \frac{4\pi a}{A} + \frac{1}{dN_1/dE} +\frac{1}{dN_2/dE}
\label{but}
\end{equation}
With the solution of $u_1$, the Eq.(\ref{u11}) becomes
\begin{eqnarray}
& & ~~ {\it region ~ I:} ~~~~-\nabla^2 u_{11} +\frac{1}{\lambda_1^2} u_{11} 
= \frac{1}{\bar{\lambda}_1^2} A_1^2 \exp(\frac{2x}{\lambda_1}) \nonumber
\\
& &~{\it region ~ II:} ~~~~ \nabla^2 u_{11}=0 \nonumber \\
& &{\it region ~ III:} ~~~~-\nabla^2 u_{11} +\frac{1}{\lambda_2^2} u_{11} 
= \frac{1}{\bar{\lambda}_2^2} A_2^2 \exp(\frac{2x}{\lambda_2})
\label{eq}
\end{eqnarray}
where we have introduced another screening length $\bar{\lambda}_\alpha^{-2} 
= 4\pi d^2\sigma_\alpha/dE^2$ corresponding to LPDOS $d^2\sigma_\alpha/dE^2$
and $A_1$ and $A_2$ are known from the calculation of $u_1$. 
The solution of Eq.(\ref{eq}) is

\begin{eqnarray}
& &~~{\it region ~ I:} ~~~~u_{11}=-\frac{\lambda_1^2}{3\bar{\lambda}_1^2}
A_1^2 \exp(\frac{2x}{\lambda_1}) + B_1 \exp(\frac{x}{\lambda_1})
\nonumber \\
& &~{\it region ~ II:} ~~~~ u_{11} = a_2 x +b_2 \nonumber \\
& &{\it region ~III:} ~~~~u_{11}=-\frac{\lambda_2^2}{3\bar{\lambda}_2^2}
A_2^2 \exp(\frac{2x}{\lambda_2}) + B_2 \exp(\frac{x}{\lambda_2})
\end{eqnarray}
After matching boundary conditions at $x=a/2,-a/2$, we obtain

\begin{equation}
B_1 \exp(\frac{-a}{2\lambda_1}) = \frac{\lambda_1}{(a+\lambda_1+\lambda_2)^3} 
[\frac{2\lambda_1^3 a +\lambda_1^4 +2\lambda_1^3
\lambda_2}{3\bar{\lambda}_1^2} + \frac{\lambda_2^4}{3\bar{\lambda}_2^2}]
\end{equation}
The second order nonlinear electrochemical capacitance $C_{111}$ is

\begin{eqnarray}
C_{111} &\equiv& \frac{1}{2} \int_{\Omega_I} \frac{\partial^2
\rho(x)}{\partial V_1^2} dx \nonumber \\
&=& \frac{-1}{4\pi} \nabla u_{11}|_{-a/2} \cdot A \nonumber \\
&=& \frac{\lambda_1}{6\pi\bar{\lambda}_1^2} A_1^2 \exp(-\frac{a}{\lambda_1})
-\frac{B_1}{4\pi \lambda_1} \exp(-\frac{a}{2\lambda_1}) \nonumber \\
&=& \frac{A}{4\pi} \frac{1}{(a+\lambda_1+\lambda_2)^3}
[\frac{\lambda_1^4}{3\bar{\lambda}_1^2} - 
\frac{\lambda_2^4}{3\bar{\lambda}_2^2}] 
\end{eqnarray}
From the definition of the screening length, we have

\begin{equation}
\frac{\lambda_\alpha}{A} = \frac{1}{4\pi A \lambda_\alpha 
d\sigma_\alpha/dE} =\frac{1}{dN_\alpha/dE}
\label{rel1}
\end{equation}
where we have used the fact that there is charge polarization only in the 
region $A\lambda_\alpha$. Similarly, we obtain

\begin{equation}
\frac{\bar{\lambda}_\alpha}{A} =\frac{1}{d^2N_\alpha/dE^2}
\label{rel2}
\end{equation}
With the help of Eqs.(\ref{rel1}), (\ref{rel2}), and (\ref{but}), we finally
have
\begin{equation}
C_{111}= \frac{C_{11}^3}{3} \left[ \frac{d^2N_1/dE^2}{(dN_1/dE)^3} - 
\frac{d^2N_2/dE^2}{(dN_2/dE)^3} \right]
\end{equation}
which agrees with Eq.(\ref{non}).

\section*{Figure Captions}

\begin{itemize}

\item[{Fig. (1)}] The energy band configuration for a model barrier.
Near the barrier the band bottom is different from that away the barrier. 
The inset is an ideal experimental setup of which the energy band is 
schematically shown in the figure.

\item[{Fig. (2)}] The linear electrochemical capacitance as a function
of barrier width $W$ for a square barrier with the barrier height $H_0=0.8$. 
The solid line is the full quantum numerical calculation $C$, the dashed line 
and dotted line are from the quantum result $C_q$ and the semi-classical 
result $C_s$ in the discrete potential approximation,
respectively. The dot-dashed line is the classical result $C\sim 1/W$. 
The inset: the corresponding partial density of states versus the barrier 
width $W$. The solid line is $dN_{11}(\Omega_{II})/dE$ and the dotted line is 
$dN_{12}(\Omega_{II})/dE$. 

\item[{Fig. (3)}] (a). The linear electrochemical capacitance as a
function of barrier width $W$ for the asymmetric barrier (see inset). 
The system parameters are $W_1=W_2$, $H_1=0.2$, $H_2=1.0$.
(b). The second order nonlinear electrochemical capacitance versus $W$. 
In Fig.(3a) and (3b), the solid, dashed, and dotted lines correspond to $C$, 
$C_q$, and $C_s$, respectively. In Fig.(3b), we have multiplied $C_q$ and 
$C_s$ by a factor of 5. The inset of Fig.(3b): the corresponding partial 
DOS $D_I$ (solid line) and $\bar{D}_I$ (dotted line) as a function of 
$W$. For illustrating purpose, we have divided $\bar{D}_I$ by a factor 
of $30$. 

\item[{Fig. (4)}] The linear (solid line) and nonlinear (dashed line) 
charge distribution for the asymmetric barrier, where $W=20$, $H_1=0.2$, 
and $H_2=1.0$. The dotted line shows the shape of asymmetric barrier.

\end{itemize}


\begin{thebibliography}{99}

\bibitem{stern}  
F. Stern, Appl. Phys. Lett. \ {\bf 43, } 974 (1983); F.
Stern, Phys. Rev. B {\bf 5}, 4891 (1972).

\bibitem{lu}  
S. Luryi, \ Appl. Phys. Lett. \ {\bf 52}(6), 501 (1988).

\bibitem{smith}  
T.P. Smith, B.B. Goldberg, P.J. Stiles and M. Heiblum, \
Phys. Rev. B \ {\bf 32}, 2696 (1985); \ \ T.P. Smith III and W.I.
Wang, \ Phys. Rev. B \ {\bf 34}, 2995 (1986).

\bibitem{butphca}  
M. B\"{u}ttiker, H. Thomas and A. Pr\^{e}tre, \ Phys.  Lett. A {\bf 180, } 
364 (1993).

\bibitem{but1}  
M. B\"{u}ttiker, \ J. \ Phys.: \ Condens. Matter \ {\bf 5,} 9361 (1993).

\bibitem{wbg1}
Due to the nonlinear bias dependence of the electrochemical capacitance,
it was proposed that a quantum scanning capacitance microscopy is 
possible.  See B.G. Wang, X.A. Zhao, J. Wang, and H. Guo, Appl. Phys. Lett. 
{\bf 74}, 2887 (1999).

\bibitem{butt2}
T. Christen and M. B\"uttiker, Phys. Rev. B {\bf 55}, R1946 (1997). 

\bibitem{chen}
W. Chen, T.P. Smith, III, M. B\"uttiker, and M. Shayegan, Phys. Rev.
Lett. {\bf 73}, 146 (1994). 

\bibitem{wei}
H.Q. Wei, N.J. Zhu, J. Wang, and H. Guo, Phys. Rev. B {\bf 56}, 9657 (1997);
P. Pomorski, H. Guo, R. Harris, and J. Wang, Phys. Rev. B {\bf 58}, 15393 
(1998). 

\bibitem{chan}
K.S. Chan et al., Phys. Rev. B {\bf 56}, 1447 (1997). 

\bibitem{wbg}
B.G. Wang, J. Wang, and H. Guo, Phys. Rev. Lett. {\bf 82}, 398 (1999).

\bibitem{jian1}
J. Wang, H. Guo, J.L. Mozos, C.C. Wan, G. Taraschi, and Q.R. Zheng,
Phys. Rev. Lett. {\bf 80}, 4277 (1998). 

\bibitem{but2}  
T. Christen and M. B\"uttiker, \ Phys. Rev. Lett. \ {\bf 77,} 143 (1996).

\bibitem{gasp}  
V. Gasparian, T. Christen, and M. B\"{u}ttiker, Phys. Rev. A \ {\bf 54}, 
4022 (1996).

\bibitem{b_euro}
T. Christen and M. B\"uttiker, Europhys. Lett. {\bf 35}  523 (1996). 

\bibitem{ma}
Z.S. Ma, J. Wang, H. Guo, Phys. Rev. B {\bf 57}, 9108 (1998);
Z.S. Ma, J. Wang, H. Guo, Phys. Rev. B {\bf 59}, 7575 (1999).

\bibitem{exp}
R.C. Ashoori et al., Phys. Rev. Lett. {\bf 68}, 3088 (1992); P. Lafarge
et al., Z. Phys. B {\bf 85}, 327 (1991); H. Tomiye and T. Yao, Appl. 
Phys. Lett. {\bf 69}, 4050 (1996); C.J. Kang et al., Appl. Phys.
Lett. {\bf 71}, 1546 (1997); J.K. Leong, C.C. Williams, J.M. Olson, Phys. 
Rev. B {\bf 56}, 1472 (1997); B. Sll, C. Gatzke, and J.M. Fernandez, 
Semicond. Sci. Technol. {\bf 13}, 423 (1998). 

\bibitem{land}  
R. Landauer, \ Philos. Mag. {\bf 21, } 863 (1970).

\bibitem{imry}  
Y. Imry, in {\it Directions in Condensed Matter Physics}, edited
by G. Grinstein and G. Mazenko, (World Scientific Singapore) pp.101, (1986).

\bibitem{buta}  
M. B\"{u}ttiker, Phys. Rev. Lett. {\bf 57}, 1761 (1986).

\bibitem{butb}  
M. B\"{u}ttiker, IBM J. Res. Develop. {\bf 32, } 317 (1988). 

\bibitem{foot4}
The physical meaning should be understood in the semiclassical limit. The
quantum value of $dn_{\alpha\alpha}/dE$ can be negative as shown in the
inset of Fig.2. For detailed discussion, see ref.\onlinecite{gasp}.

\bibitem{foot1}
Exactly the same discussion can be applied to the other two LPDOS at the
tunneling situation, $d\sigma_{11}(\Omega_{II})/dE$ and 
$d\sigma_{21}(\Omega_{II})/dE$.

\bibitem{foot2}
For nanoscale conductors with very small DOS, its electrochemical
capacitance can have a nonlinear voltage dependence due to sampling of
different parts of the DOS as the bias is varied. If this is the case, 
we should then replace Eq. (\ref{c04}) by the following equation,
\[ C_o(\frac{dU_1}{dV}-\frac{dU_2}{dV})=C_\mu+ V\frac{dC_{\mu}}{dV}. \]

\bibitem{datta} S. Datta, {\it Electronic Transport in Mesoscopic Systems},
(Cambridge University Press, New York, 1995). 

\bibitem{foot3}
Here the Thomas-Fermi approximation has been assumed. For the discussion
going beyond Thomas-Fermi approximation, see Y.D. Wei, B.G. Wang, J.
Wang, and H. Guo, cond-mat/9902306. 

\bibitem{but3}
M. B\"uttiker, H. Thomas, and A. Pr\'etre, Z. Phys. B {\bf 94}, 133
(1994). 

\bibitem{schroder}
D.K. Schroder {\it Semiconductor material and device characterization},
chapter 2, New York, John Wiley \& Sons, Inc. (1990). 

\bibitem{sohn}
L.P. Kouwenhoven et al., in {\it Mesoscopic electron transport}, edited
by L.L. Sohn et al., Dordrecht, Kluwer Academic Publishers, (1997). 

\bibitem{butt1}
T. Grammespacher and M. B\"uttiker, Phys. Rev. B {\bf 56}, 13026 (1997). 

\end{thebibliography}
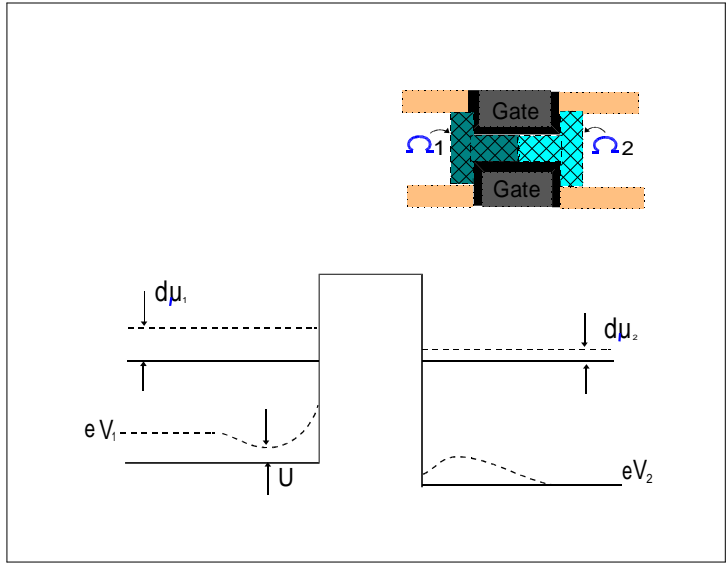
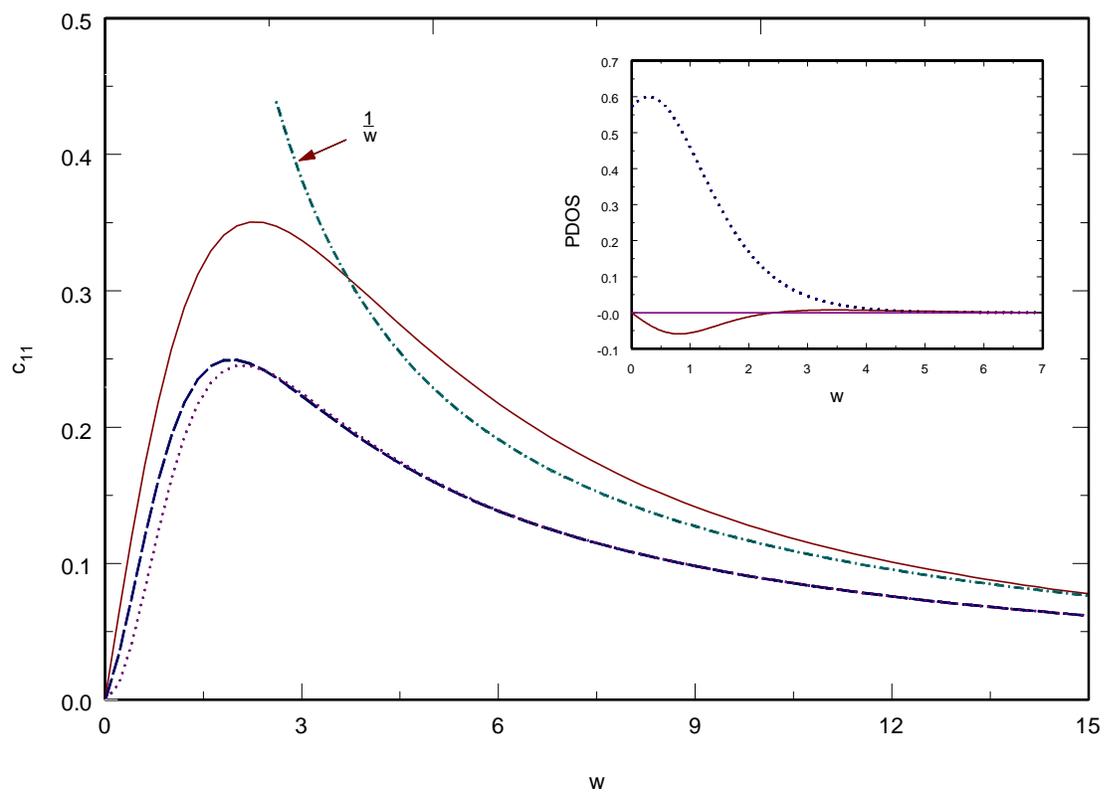
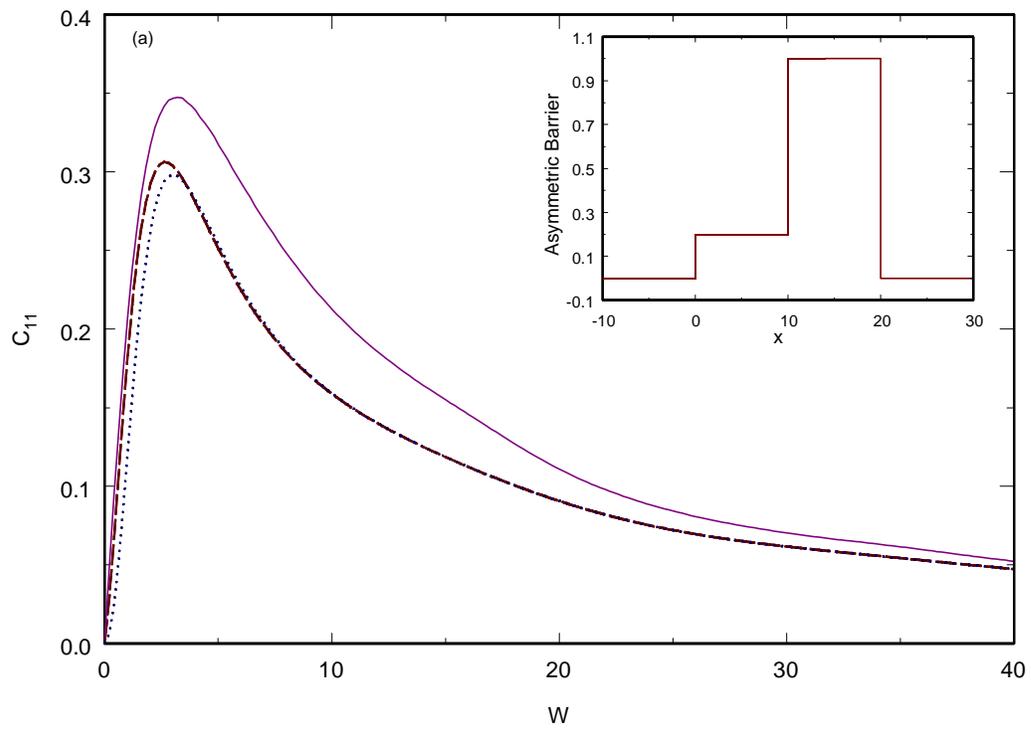
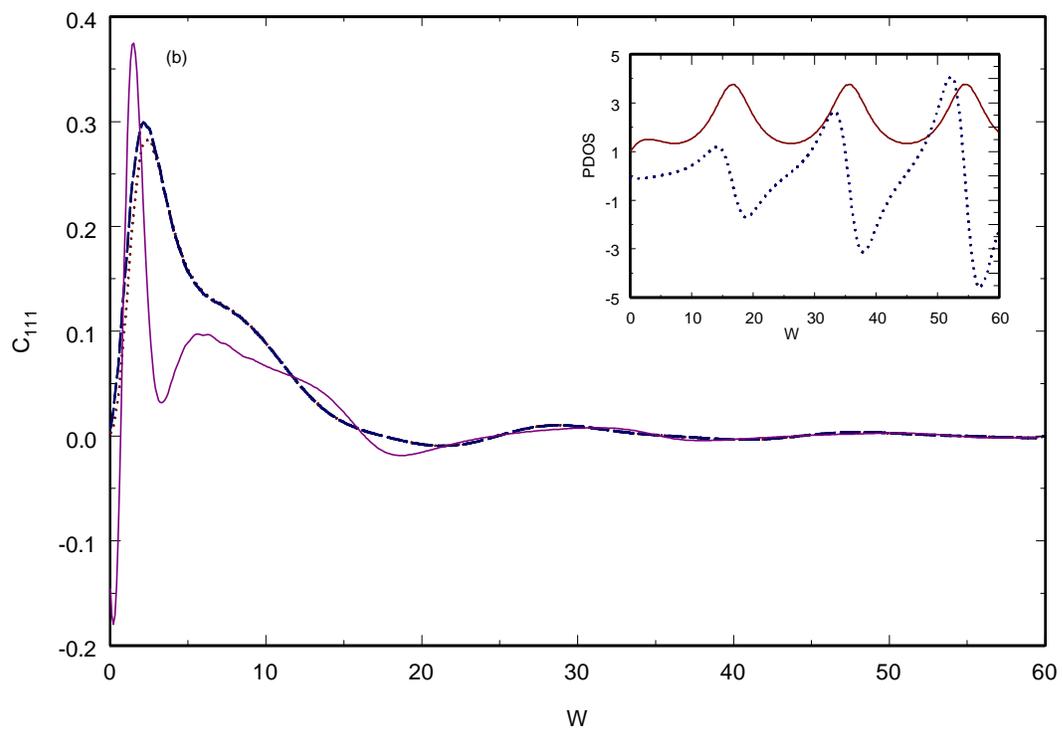
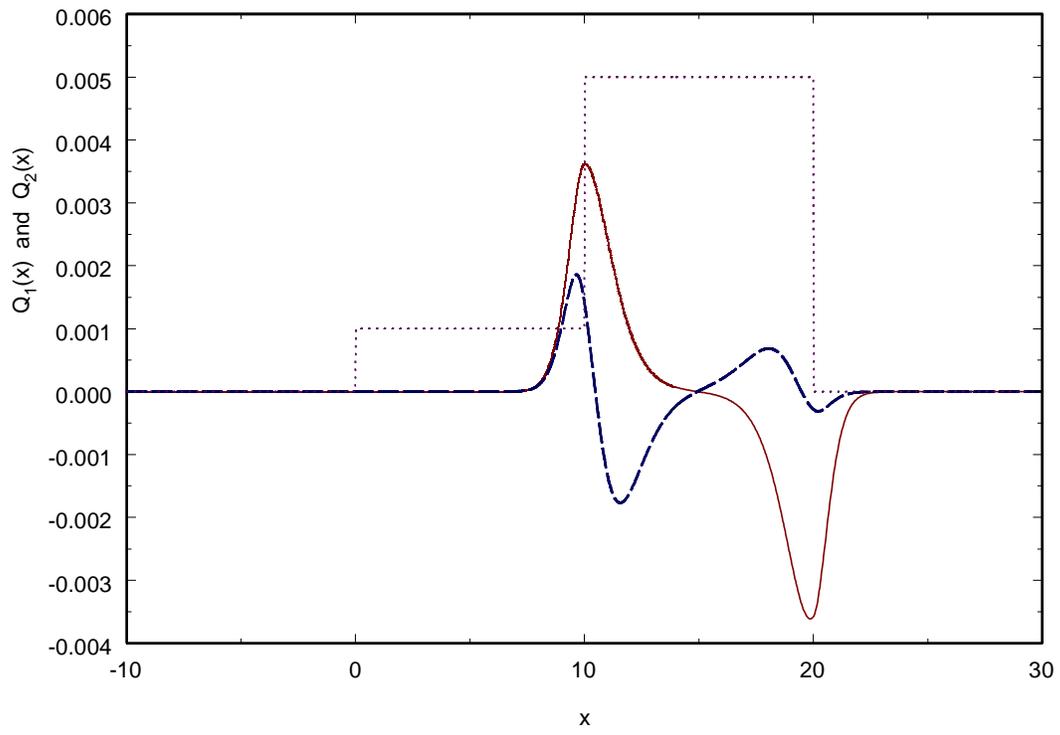
\end{document}